\documentclass[conference]{IEEEtran}
\usepackage{amsmath}
\usepackage{amsthm}
\usepackage{amsfonts}
\usepackage{amssymb}
\usepackage{authblk}
\usepackage{algorithm}
\usepackage[noend]{algpseudocode}
\usepackage{graphicx}
\usepackage{epstopdf}

\newcommand{\bx}{\mathbf{x}}
\newcommand{\bz}{\mathbf{z}}
\newcommand{\ba}{\mathbf{a}}
\newcommand{\br}{\mathbf{r}}
\newcommand{\bw}{\mathbf{w}}
\newcommand{\by}{\mathbf{y}}

\theoremstyle{definition}
\newtheorem{thm}{Theorem}

\newtheorem{exmp}[thm]{Example}

\theoremstyle{remark}
\newtheorem{defn}{Definition}

\title{Maximum Likelihood Decoder for Index Coded PSK Modulation for Priority Ordered Receivers}
\author{Divya U. S. and B. Sundar Rajan \\
Department of Electrical Communication Engineering, Indian Institute of Science, Bengaluru 560012, KA, India \\
E-mail: \{usdivya, bsrajan\}@ece.iisc.ernet.in}
\date{December 2016}

\begin{document}
\maketitle
\begin{abstract}
Index coded PSK modulation over an AWGN broadcast channel, for a given index coding problem (ICP) is studied. For a chosen index code and an arbitrary mapping (of broadcast vectors to PSK signal points), we have derived a decision rule for the maximum likelihood (ML) decoder. The message error performance of a receiver at high SNR is characterized by a parameter called \textit{\textbf{PSK Index Coding Gain (PSK-ICG)}}. The PSK-ICG of a receiver is determined by a metric called \textit{\textbf{minimum inter-set distance}}. For a given ICP with an order of priority among the receivers, and a chosen $2^N$-PSK constellation we propose an algorithm to find \textit{\textbf{(index code, mapping)}} pairs, each of which gives the best performance in terms of PSK-ICG of the receivers. No other pair of index code (of length $N$  with $2^N$ broadcast vectors) and  mapping can give a better PSK-ICG for the highest priority receiver.  Also, given that the highest priority receiver achieves its best performance, the next highest priority receiver achieves its maximum gain possible and so on in the specified order of priority.
\end{abstract}

\section{Introduction and background}
\subsection{Background}
Network coding technique has significantly improved the performance of communication networks, and has been studied extensively in the past two decades. Index coding problem (ICP) can be considered as a special case of network coding problem \cite{Ahl}. ICP has emerged as an important topic of recent research due to its applications in many of the practically relevant problems including that in satellite networks, topological interference management, wireless caching and cache enabled cloud radio access networks for 5G cellular systems. 
\par The noiseless index coding problem with side information was first studied in \cite{Birk} as an Informed-Source Coding-On-Demand (ISCOD) problem, in which a central server (sender) wants to broadcast data blocks to a set of clients (receivers) which already has a proper subset of the data blocks. The problem is to minimize the data that must be broadcast, so that each receiver can derive its required data blocks. Consider the case of a sender with $n$ messages denoted by the set $\mathcal{X} = \{x_1,x_2,...,x_n\}$, $x_i ~\in~ \mathbb{F}_{q}$, $\mathbb{F}_{q}$ is a field with $q$ elements, which it broadcasts as coded messages, to a set of $m$ receivers, $\mathcal{R}=\{R_1,R_2,...,R_m\}$. Each receiver $R_i ~\in~\mathcal{R}$ wants a subset $\mathcal{W}_{i}$ of the messages, knows a priori a proper subset $\mathcal{K}_{i}$ of the messages, where $\mathcal{W}_{i}~\cap~\mathcal{K}_{i}=\phi$, and is identified by the pair $(\mathcal{W}_{i},\mathcal{K}_{i})$. The noiseless index coding problem is to find the smallest number of transmissions required and is specified by $(\mathcal{X},\mathcal{R})$. The set $\mathcal{K}_{i}$ is referred to as the side information available to the receiver $R_i$. 
\begin{defn} 
An index code (IC) for a given ICP $(\mathcal{X},\mathcal{R})$ is defined by an encoding function,
 $g: \mathbb{F}^{n}_{q} \rightarrow \mathbb{F}^{l}_{q}, $ 
and a set of $m$ decoding functions 
$\mathcal{D}_i:  \mathbb{F}^{l}_{q} \times \mathbb{F}^{|\mathcal{K}_{i}|}_{q} \rightarrow \mathbb{F}^{|\mathcal{W}_{i}|}_{q}, $ $\forall~ i~ \in~ \{1,2,...,m\}$
 corresponding to the $m$ receivers, such that,
\begin{equation*}
\mathcal{D}_{i}(g(\textbf{x}),\mathcal{K}_{i}) = \mathcal{W}_{i},~~~\forall~ \textbf{x}~ \in~ \mathbb{F}^{n}_{q},~~~\forall~i~ \in~ \{1,2,...,m\}.
\end{equation*}
\end{defn}

\par In this paper, we consider ICP over binary field ($q=2$). The integer $l$, as defined above is called the length of the index code. For noiseless broadcast channels, an index code of minimum length is called an optimal index code \cite{Ong}, \cite{Yossef}.  
Even though this is interesting theoretically, index coding over noisy channels is more practical. Noisy index coding over a binary symmetric channel was considered in \cite{Anp1}, \cite{Anp2}. Binary transmission of index coded bits were assumed. 
For this set up the problem of identifying the number of optimal index codes possible for a given ICP is important and that was studied in \cite{Kav1}, \cite{Kav2}.
\par A special case of ICP over Gaussian broadcast channel, based on multidimensional QAM constellation with $2^n$ points, where every receiver demands all messages (which it does not have) from the source, was considered in \cite{Lpn}. 
The case of noisy index coding over AWGN broadcast channel, along with minimum Euclidean distance decoding, was studied in \cite{Anj1}, where the receivers demand a subset of messages as defined in \cite{Birk}. 
An algorithm to map the broadcast vectors to PSK signal points so that the receiver with maximum side information gets maximum PSK side information coding gain, was also proposed. The algorithm assumes that an index code is given and is applicable only for one specific order of priority (in the non increasing order of amount of side information) among the receivers. Minimum Euclidean distance of the effective broadcast signal set seen by a receiver, was considered as the basic parameter which decides the message error probability of the receiver and the proposed algorithm tries to maximize the minimum Euclidean distance.
\par In this paper, we discuss the maximum likelihood (ML) decoder for index coded PSK modulation. Further, we study the case in which the length of the index code is specified for the ICP but not necessarily the index code. The receivers can have any a priori defined arbitrary order of priority among themselves. For a chosen priority order,  we consider all possible index codes, to obtain the mappings to appropriate PSK constellation which will result in the best message error performance in terms of PSK index coding gain (PSK-ICG, defined in Section \ref{isd_section}) of the receivers, respecting the defined order of priority.

\subsection{Our Contribution}
Consider a noisy index coding problem with $n$ messages, over $\mathbb{F}_2$ which uses an AWGN broadcast channel for transmission. For the ICP $(\mathcal{X},\mathcal{R})$, consider index codes of length $N$, $N<n$, which will generate $2^N$ broadcast vectors (elements of $\mathbb{F}^N_2$). The broadcast vectors are mapped to $2^N$-PSK signal points, so that $2^N$-PSK modulation can be used, to minimize the bandwidth requirement. Note that, transmitting one $2^N$-PSK signal point instead of $N$ binary bits (as in noiseless index coding), results in $N/2$ fold saving in bandwidth.
\par Our contributions are summarized below: 

\begin{itemize}
\item We derive a decision rule for maximum likelihood decoding which gives the best message error performance, for any receiver $R_i$, for a given index code and mapping.
\item We show that, at very high SNR, the message error performance of the receiver employing ML decoder, depends on the minimum inter-set distance (defined in Section \ref{isd_section}). The mapping which maximized the minimum inter-set distance is optimal for the best message error performance at high SNR.
\item For the ICP $(\mathcal{X},\mathcal{R})$, when the receivers are arranged in the decreasing order of priority, we propose an algorithm to find (index code, mapping) pairs, each of which gives the best message error performance for the receivers, for the given order of priority. Using any one of the above (index code, mapping) pairs, the highest priority receiver achieves the maximum possible gain (PSK-ICG) that it can get using any IC and any mapping for $2^N$-PSK constellation, at very high SNR. Given that the highest priority receiver achieves its best performance, the next highest priority receiver achieves its maximum gain possible  and so on in the specified order of priority.
\end{itemize}

\section{Preliminaries and notation}
\par Let $[n] \triangleq \{1,2,...,n\}$. For a vector $\bz=(z_1~z_2~...z_n) \in \mathbb{F}^n_{2}$ and a subset $B=\{i_1,i_2,...,i_b\}$ of $[n]$ (for any integer $b, 1 \leq b \leq n$), where $i_1<i_2<...<i_b$, $\bz_B$
denote the vector $(z_{i_1}~z_{i_2}~...~z_{i_b})$. 
 \par We consider the noisy index coding problem over $\mathbb{F}_2$ with a single sender having a set of messages $\mathcal{X} = \{x_1,x_2,...,x_n\}$, $x_i\in\mathbb{F}_{2}$, and a set of $m$ receivers, $\mathcal{R}=\{R_1,R_2,...,R_m\}$, where each receiver $R_i$ is identified by $(\mathcal{W}_{i},\mathcal{K}_{i})$, the want set and the known set. Let, $\mathcal{I}_{i} \triangleq \{j : x_j \in \mathcal{K}_i\}$ be the set of indices corresponding to the known set. It is sufficient to consider the case where each receiver demands only one message. If there is a receiver which demands more than one message, it can be considered as $|\mathcal{W}_{i}|$ equivalent receivers each demanding one message and having the same side information. Each $R_i$, $i \in [m]$ wants the message $x_{f(i)}$, where $f:[m] \rightarrow [n]$ and $x_{f(i)} \notin \mathcal{K}_{i}$, $\forall i \in [m]$. 
\par For the given ICP, we consider scalar linear index codes of length $N$ (not necessarily the minimum or optimum length),  such that the set of all broadcast vectors gives $\mathbb{F}_2^N.$ Let $L$ be an $n~\times~N$ encoding matrix for one such index code, $\mathcal{C}$. Let $\bx=(x_1~x_2~ ...~x_n)$ and $\by =(y_1~y_2~...y_N)$ denote the message vector and the broadcast vector respectively, where $\by = \bx L.$ 

\begin{exmp} \label{exmp:eg1}
Consider the following ICP with $n=m=5$ and $\mathcal{W}_{i}=x_i, ~\forall i \in \{1,2,...,5\}$. The side information available with the receivers is as follows: $\mathcal{K}_1= \{x_2, x_3\}, ~\mathcal{K}_2= \{x_3, x_4, x_5\},~ \mathcal{K}_3= \{x_2, x_4, x_5\},~\mathcal{K}_4= \{x_5\},~\mathcal{K}_5= \{x_4\}$.
\par For this ICP we can choose a scalar linear index code of length $N=3$, as given by the following encoding matrix $L$.
	\[
	L=
	\begin{bmatrix}
	1 & 1 & 0 \\
	0 & 1 & 0 \\
	0 & 1 & 0 \\
	1 & 1 & 1 \\
	1 & 1 & 1 
	\end{bmatrix}
	\]. \\
The index coded bits are given by \\
$(y_1~y_2~y_3)=(x_1~x_2~x_3~x_4~x_5)L$ as $y_1=x_1+x_4+x_5,~ y_2=x_1+x_2+x_3+x_4+x_5,~ y_3=x_4+x_5$.
\end{exmp}

\par Instead of using $N$ BPSK transmissions, the $N$ index coded bits of $\by$ are sent as a signal point from a $2^N$-PSK signal set, over an AWGN channel, to save bandwidth \cite{Anj1}. In this paper we consider index coded $2^N$-PSK modulation for a chosen $N$ and so when we refer to index codes of length $N$, we consider only those index codes for which the set of all broadcast vectors is $\mathbb{F}^N_{2}$. Let the chosen $2^N$-PSK signal set be denoted as $\mathcal{S}=\{\mathbf{s_{1}},\mathbf{s_{2}},...,\mathbf{s_{2^N}}\}$. Assume that for the index code $\mathcal{C}$ a mapping scheme specifies the mapping of $\mathbb{F}^N_{2}$ to the signal set $\mathcal{S}$. All receivers are assumed to know the encoding matrix, $L$ for the index code $\mathcal{C}$. 
\par Let $\ba_i \in \mathbb{F}^{|\mathcal{K}_i|}_{2}$ be a realization of $\bx_{\mathcal{I}_{i}}$. As each receiver $R_i$ knows some messages (from its side information), $R_i$ needs to consider only a subset of $\mathbb{F}^N_{2}$ for decoding and this subset is called the {\it effective broadcast vector set}. 

\begin{defn}
For a chosen index code based on the encoding matrix $L$, the {\it effective broadcast vector set} seen by $R_i$ for $\bx_{\mathcal{I}_{i}}=\ba_i$ is defined by,
\begin{equation*}
\begin{aligned}
\mathcal{C}_L(\ba_i) &\triangleq \{\by\in\mathbb{F}^N_2 : \by= \bx L, \bx_{\mathcal{I}_{i}}=\ba_i, x_j \in \mathbb{F}_2, j \in [n] \setminus \mathcal{I}_{i}\}.
\end{aligned}
\end{equation*}
\end{defn}

The corresponding set of signal points in $2^N$-PSK constellation is referred to as the {\it effective broadcast signal set} seen by $R_i$ for $\bx_{\mathcal{I}_{i}}=\ba_i$ and is denoted by $\mathcal{S}_L(\ba_i)$. For a chosen index code, all effective broadcast signal sets and effective broadcast vector sets seen by $R_i$ are of the same size ($|\mathcal{S}_L(\ba_i)|=|\mathcal{S}_L(\ba'_i)|=|\mathcal{C}_L(\ba_i)|=|\mathcal{C}_L(\ba'_i)|$ where $\ba_i,\ba'_i \in \mathbb{F}^{|\mathcal{K}_i|}_{2}$).

Half the number of broadcast vectors in an effective broadcast vector set corresponds to $x_{f(i)}=0$ and the remaining half corresponds to $x_{f(i)}=1$. So, we can partition an effective broadcast vector set into two subsets as defined below.

\begin{defn}
The {\it 0-effective broadcast vector set} seen by $R_i$ for $\bx_{\mathcal{I}_{i}}=\ba_i$ is defined by,
\begin{equation*}
\begin{aligned}
\mathcal{C}_{L0}(\ba_i) \triangleq \{ &\by~\in~\mathbb{F}^N_2 : \by= \bx L, \bx_{\mathcal{I}_{i}}=\ba_i, x_{f(i)}=0, x_j \in \mathbb{F}_2, \\ &j \in [n] \setminus (\mathcal{I}_{i} \cup \{f(i)\})\}.
\end{aligned}
\end{equation*}
\end{defn}
The corresponding set of signal points in $2^N$-PSK constellation is referred to as the {\it 0-effective broadcast signal set} seen by $R_i$ for $\bx_{\mathcal{I}_{i}}=\ba_i$ and is denoted as $\mathcal{S}_{L0}(\ba_i)$.

\begin{defn}
The {\it 1-effective broadcast vector set} seen by $R_i$ for $\bx_{\mathcal{I}_{i}}=\ba_i$ is defined by,
\begin{equation*}
\begin{aligned}
\mathcal{C}_{L1}(\ba_i) \triangleq \{ &\by~\in~\mathbb{F}^N_2 : \by= \bx L, \bx_{\mathcal{I}_{i}}=\ba_i, x_{f(i)}=1, x_j \in \mathbb{F}_2, \\ &j \in [n] \setminus (\mathcal{I}_{i} \cup \{f(i)\})\}.
\end{aligned}
\end{equation*}
\end{defn}
The corresponding set of signal points in $2^N$-PSK constellation is referred to as the {\it 1-effective broadcast signal set} seen by $R_i$ for $\bx_{\mathcal{I}_{i}}=\ba_i$ and is denoted as $\mathcal{S}_{L1}(\ba_i)$.

The effective broadcast vector sets, 0-effective broadcast vector sets and 1-effective broadcast vector sets seen by $R_2$ for the IC in Example \ref{exmp:eg1} is given in Table \ref{table:tbl1}. It is clear that, two different realizations of $\bx_{\mathcal{I}_{i}}$ may have the same effective broadcast vector set. However, 1-effective broadcast vector set for a particular realization of $\bx_{\mathcal{I}_{i}}$ may become the 0-effective broadcast vector set of another realization of $\bx_{\mathcal{I}_{i}}$ and vice versa. But the way in which the effective broadcast vector set gets partitioned will be the same. For example consider the case of $\mathcal{C}_L(011)$ and $\mathcal{C}_L(100)$ in Table \ref{table:tbl1}.

\begin{table}[h]
\caption{Effective broadcast vector sets and its partitions (seen by $R_2$) for the IC in Example \ref{exmp:eg1}.}
\centering
\begin{tabular}{p{0.5cm}|p{3.4cm}|p{1.6cm}|p{1.6cm}}
\hline
\hline                     
~~$\ba_2$ & ~~~~~~~~~~~$\mathcal{C}_L(\ba_2)$ & ~~~~$\mathcal{C}_{L0}(\ba_2)$ & ~~~~$\mathcal{C}_{L1}(\ba_2)$\\
\hline
$(000)$ & $\{(000),(010),(110),(100)\}$ & $\{(000),(110)\}$ & $\{(010),(100)\}$ \\
$(001)$ & $\{(111),(101),(001),(011)\}$ & $\{(111),(001)\}$ & $\{(101),(011)\}$  \\
$(010)$ & $\{(111),(101),(001),(011)\}$ & $\{(111),(001)\}$ & $\{(101),(011)\}$  \\
$(011)$ & $\{(000),(010),(110),(100)\}$ & $\{(000),(110)\}$ & $\{(010),(100)\}$ \\
$(100)$ & $\{(000),(010),(110),(100)\}$ & $\{(010),(100)\}$ & $\{(000),(110)\}$ \\
$(101)$ & $\{(111),(101),(001),(011)\}$ & $\{(101),(011)\}$ & $\{(111),(001)\}$  \\
$(110)$ & $\{(111),(101),(001),(011)\}$ & $\{(101),(011)\}$ & $\{(111),(001)\}$  \\
$(111)$ & $\{(000),(010),(110),(100)\}$ & $\{(010),(100)\}$ & $\{(000),(110)\}$ \\ [1ex]      
\hline
\end{tabular}
\label{table:tbl1}
\end{table}

\begin{exmp} \label{exmp:eg2}
Consider the following ICP with $n=m=6$ and $\mathcal{W}_{i}=x_i, ~\forall i \in \{1,2,...,6\}$. The side information available with the receivers is as follows: $\mathcal{K}_1= \{x_2, x_3, x_4, x_5, x_6\}, ~\mathcal{K}_2= \{x_1, x_3, x_4, x_5\},~ \mathcal{K}_3= \{x_2, x_4, x_6\},~\mathcal{K}_4= \{x_1, x_6\},~\mathcal{K}_5= \{x_3\},~\mathcal{K}_6= \{\} $.
\par For this ICP we can choose a scalar linear index code of length $N=4$, based on encoding matrix $L$, with $y_1=x_1+x_4,~ y_2=x_2+x_3,~ y_3=x_5,~y_4=x_6$. \\
Then, the effective broadcast vector sets of $R_2$ for four different realization of $\bx_{\mathcal{I}_{2}}$ are as given below. 
\begin{itemize}
\item $\mathcal{C}_L(0000)=\{(0000),(0100),(0001),(0101)\}$
\item $\mathcal{C}_L(0001)=\{(0010),(0110),(0011),(0111)\}$
\item $\mathcal{C}_L(0010)=\{(1000),(1100),(1001),(1101)\}$ 
\item $\mathcal{C}_L(0011)=\{(1010),(1110),(1011),(1111)\}$.
\end{itemize}
\end{exmp}

Suppose an IC based on an encoding matrix $L$, and an effective broadcast vector set, $\mathcal{C}_L(\ba_i)$ of $R_i$ are given. $\mathcal{C}_L(\ba_i)$ can be partitioned into 0-effective broadcast vector set and 1-effective broadcast vector set as follows:
\begin{itemize}
\item Identify an $\bx$ such that $\bx L \in \mathcal{C}_L(\ba_i)$. Let the corresponding realization of $\bx_{\mathcal{I}_{i}}$ be $\ba_i$ .
\item For $\ba_i$, partition $\mathcal{C}_L(\ba_i)$ into $\mathcal{C}_{L0}(\ba_i)$ and $\mathcal{C}_{L1}(\ba_i)$
\end{itemize}
The partitioning of $\mathcal{C}_L(\ba_i)$ is illustrated in the following example. Consider the ICP given in Example \ref{exmp:eg2}. Suppose the effective broadcast vector set, $\mathcal{C}_L(\ba_2)=\{(0000),(0100),(0001),(0101)\}$ of $R_2$ needs to be partitioned into $\mathcal{C}_{L0}(\ba_2)$ and $\mathcal{C}_{L1}(\ba_2)$. Choose $\bx=(110100)$ such that $\by= \bx L=(0100) \in \mathcal{C}_L(\ba_2)$, and then $\ba_2=(1010)$. Note that $y_2=x_2+x_3$, $x_3 \in \mathcal{K}_{2}$, $R_2$ wants $x_2$, and from $\ba_2=(1010)$, $x_3=0$. So $y_2=x_2$ and only two broadcast vectors, $(0000)$ and $(0001)$ in $\mathcal{C}_L(1010)$ has $y_2=0$. So $\mathcal{C}_{L0}(1010)=\{(0000),(0001)\}$. Similarly, $\mathcal{C}_{L1}(1010)=\{(0100),(0101)\}$. It should be noted that for some other choice of $\bx$ with $x_3=1$, we may get $\mathcal{C}_{L0}(\ba_2)=\{(0100),(0101)\}$ and $\mathcal{C}_{L1}(\ba_2)=\{(0000),(0001)\}$. We are only interested in partitioning the effective broadcast vector set into two subsets such that all broadcast vectors in each subset correspond to the same value of $x_{f(i)}$.

\begin{figure*}
\begin{center}
\includegraphics[scale=1]{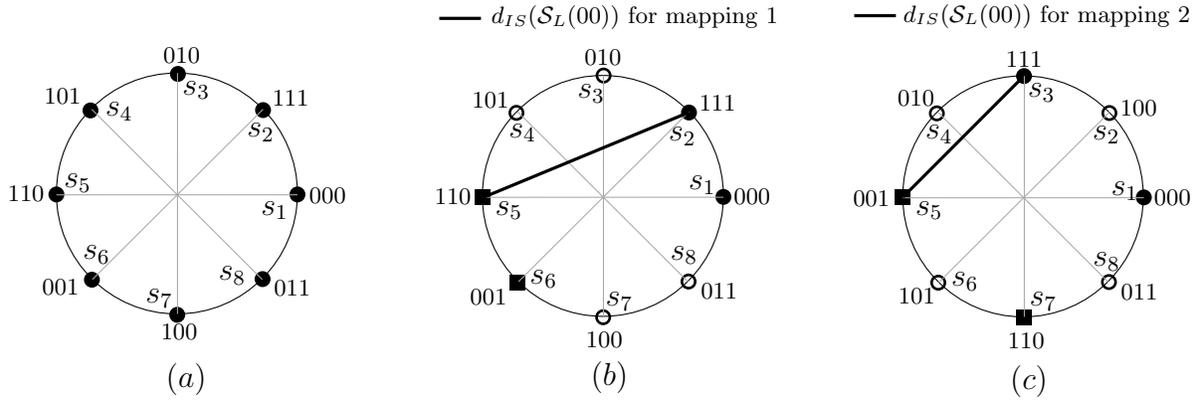} 
\caption{8-PSK mapping and inter-set distance for $R_1$ in Example \ref{exmp:eg1}.}
\label{figure:fig1}
\end{center}
\end{figure*}

\section{Maximum likelihood decoder}
In this section we derive a decision rule for the maximum likelihood decoder for the receiver $R_i$. We follow an approach similar to the one used in \cite{Vit}. 
\par Let $\mathcal{M}$ be the map from $\mathbb{F}^N_{2}$ to the signal set $\mathcal{S}$. The received vector $\br$ is given by
\begin{equation*}
\br=\mathcal{M}(\bx L)+\bw
\end{equation*}
where $\bw=(w_1 ~w_2)$; $w_1$ and $w_2$ are independent Gaussian variables with zero mean and variance $N_0/2$. The conditional probability density of $\br$ given that $\mathcal{M}(\bx L)$ is transmitted (likelihood function) is
\begin{multline}
p(\br|\mathcal{M}(\bx L)) = \frac{1}{(\pi N_0)}\exp{\left(-\frac{\|\br-\mathcal{M}(\bx L)\|^2}{N_0}\right)}. \label{eqawgn}
\end{multline}

Consider the decoder for a receiver $R_i$. The minimum error probability decoder should make a decision $x'_{f(i)}$ on the desired message $x_{f(i)}$ based on the received vector $\br$ and the side information $\bx_{\mathcal{I}_{i}}$, minimizing the probability of error. Given $\bx_{\mathcal{I}_{i}}=\ba_i$, when $x_{f(i)}=0$ the probability of error in this decision is $\textbf{Pr}(x_{f(i)}=1|\bx_{\mathcal{I}_{i}}=\ba_i,\br)$ and that when $x_{f(i)}=1$ is $\textbf{Pr}(x_{f(i)}=0|\bx_{\mathcal{I}_{i}}=\ba_i,\br)$.
To minimize the error probability, the decision $x'_{f(i)}=0$ is taken if
\begin{equation}
\begin{aligned}
\textbf{Pr}(x_{f(i)}=0|\bx_{\mathcal{I}_{i}}=\ba_i,\br) \geq \textbf{Pr}(x_{f(i)}=1|\bx_{\mathcal{I}_{i}}=\ba_i,\br) \label{eq1}
\end{aligned}
\end{equation}
and the decision $x'_{f(i)}=1$ is taken if
\begin{equation}
\begin{aligned}
\textbf{Pr}(x_{f(i)}=0|\bx_{\mathcal{I}_{i}}=\ba_i,\br) < \textbf{Pr}(x_{f(i)}=1|\bx_{\mathcal{I}_{i}}=\ba_i,\br). \label{eq2}
\end{aligned}
\end{equation}
Combining (\ref{eq1}) and (\ref{eq2}), and ignoring ties, the decision rule can be written as
\begin{equation}
\begin{aligned}
\textbf{Pr}(x_{f(i)}=0|\bx_{\mathcal{I}_{i}}=\ba_i,\br) \underset{0}{\overset{1}\lessgtr}   \textbf{Pr}(x_{f(i)}=1|\bx_{\mathcal{I}_{i}}=\ba_i,\br). \label{eq3}
\end{aligned}
\end{equation}
Using Bayes rule in (\ref{eq3}), we obtain the decision rule in terms of the likelihood functions as
\begin{multline*}
\frac{p(\br|x_{f(i)}=0,\bx_{\mathcal{I}_{i}}=\ba_i)\textbf{Pr}(x_{f(i)}=0)}{p(\br)} \underset{0}{\overset{1}\lessgtr}  \\ \frac{p(\br|x_{f(i)}=1,\bx_{\mathcal{I}_{i}}=\ba_i)\textbf{Pr}(x_{f(i)}=1)}{p(\br)}, 
\end{multline*}
which implies
\begin{multline}
p(\br|x_{f(i)}=0,\bx_{\mathcal{I}_{i}}=\ba_i)\textbf{Pr}(x_{f(i)}=0) \underset{0}{\overset{1}\lessgtr}  \\ 
p(\br|x_{f(i)}=1,\bx_{\mathcal{I}_{i}}=\ba_i)\textbf{Pr}(x_{f(i)}=1). \label{eq5}
\end{multline}
$\mathcal{S}_{L0}(\ba_i)$, the 0-effective broadcast signal set seen by $R_i$ (for $\ba_i$), is the set of all signal points corresponding to broadcast vectors with $x_{f(i)}=0$ and $\bx_{\mathcal{I}_{i}}=\ba_i$. Therefore, 
\begin{equation}
\begin{aligned}
p(\br|x_{f(i)}=0,\bx_{\mathcal{I}_{i}}=\ba_i) &= p(\br|\mathcal{S}_{L0}(\ba_i)). \label{eqeff0}
\end{aligned}
\end{equation}
Similarly,
\begin{equation}
\begin{aligned}
p(\br|x_{f(i)}=1,\bx_{\mathcal{I}_{i}}=\ba_i) &= p(\br|\mathcal{S}_{L1}(\ba_i)). \label{eqeff1}
\end{aligned}
\end{equation}
Assuming that all the messages take values 0 or 1 with equal probability, from (\ref{eq5}), (\ref{eqeff0}) and (\ref{eqeff1}) we obtain the decision rule as
\begin{equation}
 \sum_{k:s_k \in \mathcal{S}_{L0}(\ba_i) } p(\br|s_k) \underset{0}{\overset{1}\lessgtr}   \sum_{k:s_k \in \mathcal{S}_{L1}(\ba_i) } p(\br|s_k).  \label{eqml2}
\end{equation}
From (\ref{eqawgn}) and (\ref{eqml2}),
\begin{multline*}
\sum_{k:s_k \in \mathcal{S}_{L0}(\ba_i)} {\left(\frac{1}{(\pi N_0)}\exp{\left(-\frac{\|\br-s_k\|^2}{N_0}\right)}\right)} 
 \underset{0}{\overset{1}\lessgtr}  \\
\sum_{k:s_k \in \mathcal{S}_{L1}(\ba_i)} {\left(\frac{1}{(\pi N_0)}\exp{\left(-\frac{\|\br-s_k\|^2}{N_0}\right)}\right)}. 
\end{multline*}
Thus we obtain the ML decision rule as,
\begin{multline}
\sum_{k:s_k \in \mathcal{S}_{L0}(\ba_i)}{\left( \exp{\left(-\frac{\|\br-s_k\|^2}{N_0}\right)}\right)}
\underset{0}{\overset{1}\lessgtr}  \\
\sum_{k:s_k \in \mathcal{S}_{L1}(\ba_i)}{\left( \exp{\left(-\frac{\|\br-s_k\|^2}{N_0}\right)}\right)}. \label{eqml3}
\end{multline}
It is clear that the ML decoder decision is based on the Euclidean distance of all signal points in 0-effective broadcast signal set to the received vector $\br$ relative to that of the signal points in 1-effective broadcast signal set. This indicates that, to reduce the message error probability, the signal points in 0-effective broadcast signal set and 1-effective broadcast signal set must be as separated as possible in terms of Euclidean distance.

\section{Inter-set distance and PSK Index Coding Gain} \label{isd_section}

\begin{defn} 
{\it Inter-set distance} of an effective broadcast signal set seen by a receiver $R_i$ is the minimum among the Euclidean distances between a signal point in the 0-effective broadcast signal set and a signal point in the 1-effective broadcast signal set.
\begin{equation*}
\begin{aligned}
d_{IS}(\mathcal{S}_L(\ba_i)) \triangleq \min\{d(\mathbf{s_a},\mathbf{s_b}):&\mathbf{s_a} \in \mathcal{S}_{L0}(\ba_i), \mathbf{s_b} \in \mathcal{S}_{L1}(\ba_i)\}
\end{aligned}
\end{equation*}
where $d(\mathbf{s_a},\mathbf{s_b})$ denotes the Euclidean distance between $2^N$-PSK signal points, $\mathbf{s_a}$ and $\mathbf{s_b}$.
\end{defn}
A labeled 8-PSK constellation which can be used for the ICP discussed in Example \ref{exmp:eg1} is shown in Fig. \ref{figure:fig1}(a) and the inter-set distance of the effective broadcast signal set seen by $R_1$ for $\bx_{\mathcal{I}_{i}}=(00)$, $\mathcal{S}_L(00)$ is shown in Fig. \ref{figure:fig1}(b). For this example, $\mathcal{S}_L(00)=\{\mathbf{s_1},\mathbf{s_2},\mathbf{s_5},\mathbf{s_{6}}\}$, $\mathcal{S}_{L0}(00)=\{\mathbf{s_1},\mathbf{s_{2}}\}$ and $\mathcal{S}_{L1}(00)=\{\mathbf{s_5},\mathbf{s_6}\}$.

\begin{defn} 
For a given index code and mapping, the minimum inter-set distance for a receiver $R_i$, denoted by $d^{(i)}_{IS,min}$, is defined as the minimum of the inter-set distances among all the effective broadcast signal sets seen by $R_i$.
\begin{equation*}
d^{(i)}_{IS,min} \triangleq \min\{d_{IS}(\mathcal{S}_L(\ba_i)):\ba_i~\in~\mathbb{F}^{|\mathcal{K}_i|}_2\}
\end{equation*}
\end{defn}
In the case of Example \ref{exmp:eg1}, the minimum inter-set distance for $R_1$ is shown in Fig. \ref{figure:fig1}(b) and \ref{figure:fig1}(c) for two different mappings. Clearly, the mapping shown in Fig. \ref{figure:fig1}(b) has a larger minimum inter-set distance for $R_1$.

\par In (\ref{eqml3}), the term with the signal point closest to $\br$ is dominant in the summations at very high SNR. The decoder makes an error if the broadcasted signal point is in 0-effective broadcast signal set but $\br$ is closest to a signal point in 1-effective broadcast signal set or vice versa. The probability of this event is more when the minimum inter-set distance is less. At high SNR, this error is dominant and so an optimal mapping for the best message error performance must maximize the minimum inter-set distance. Among the mappings which has the same minimum inter-set distance, the one which has more second minimum inter-set distance will perform better and so on.

\begin{defn} 
The PSK Index Coding Gain (PSK-ICG) of a receiver $R_i$, for a given IC and mapping is defined as
\begin{equation*}
g_{i} \triangleq 20~log\left(\frac{d^{(i)}_{IS,min}}{d_{min,n}}\right)
\end{equation*}
where $d^{(i)}_{IS,min}$ is the minimum inter-set distance for $R_i$ and $d_{min,n}$ is the minimum Euclidean distance between any two signal points in a $2^n$-PSK constellation.
\end{defn}

\section{Mapping based on inter-set distances}
For mapping, it is more appropriate to consider the minimum inter-set distance than to consider the minimum Euclidean distance of the effective broadcast signal sets. For example, consider the mappings given in Fig. \ref{figure:fig1}(b) and Fig. \ref{figure:fig1}(c). With the mapping shown in Fig. \ref{figure:fig1}(b), the minimum inter-set distance for $R_1$ is more but the minimum Euclidean distance of its effective broadcast signal sets is less, compared to that with the mapping shown in Fig. \ref{figure:fig1}(c). The simulation results (discussed in Section \ref{sim_section}) show that $R_1$ performs better with the mapping in Fig. \ref{figure:fig1}(b) than with the mapping in Fig. \ref{figure:fig1}(c).
\par For the given ICP and $2^N$-PSK constellation, when the receivers are arranged in the decreasing order of priority, we propose an algorithm which maximizes the minimum inter-set distance, to find (index code, mapping) pairs, each of which gives the optimal message error performance for the receivers, for the given order of priority. Assume that the decreasing order of priority for the receivers is $(R_1,R_2,...,R_m)$. Here optimality is based on minimum inter-set distance and is in the following sense:\\

\begin{itemize}
\item No other mapping of $2^N$-PSK constellation for any index code, can give PSK-ICG $>g_{1}$ for $R_1$.
\item Any mapping for any index code which gives the PSK-ICG $g_i$ for receiver $R_i$, $i \in \{1,2,...,j-1\}$ cannot give a PSK-ICG  $>g_j$ for $R_j$, $j \leq m$.
\end{itemize}

It may so turn out that maximizing the gain of a receiver $R_i$, minimizes the gain that can be achieved by a lower priority receiver $R_j$. With this mapping it is not necessary that a higher priority receiver will get higher PSK-ICG compared to that of the lower priority receivers. The PSK-ICG achieved by a receiver $R_j$ depends on its priority, $\mathcal{W}_j$, $\mathcal{K}_j$, $\mathcal{W}_i$ and $\mathcal{K}_i$ $\forall i$ such that $R_i$ is a higher priority receiver than $R_j$. 

In the following subsections, we explain the mapping algorithm and then illustrate it with examples.

\subsection{Mapping Algorithm}
Without loss of generality, assume that the decreasing order of priority among the receivers is $(R_1,R_2,...,R_m)$.  
For a given index code based on encoding matrix $L$, optimal mapping for a receiver $R_i$ is obtained as follows:
\begin{enumerate}
\item Find all effective broadcast vector sets for $\ba_i \in \mathbb{F}^{|\mathcal{K}_i|}_{2}$ . These sets partition $\mathbb{F}_2^N$.  
\item Consider an effective broadcast vector set, $\mathcal{C}_L(\ba_i)$.  
\item \label{map-begn} Partition the effective broadcast vector set into 0-effective broadcast vector set ($\mathcal{C}_{L0}(\ba_i)$) and 1-effective broadcast vector set ($\mathcal{C}_{L1}(\ba_i)$).
\item All the broadcast vectors in $\mathcal{C}_{L0}(\ba_i)$ must be mapped to adjacent signal points. Let the set of signal points corresponding to $\mathcal{C}_{L0}(\ba_i)$ be $\mathcal{S}_{L0}(\ba_i)$.
\item \label{map-end} All the broadcast vectors in $\mathcal{C}_{L1}(\ba_i)$ must be mapped to signal points diametrically opposite to signal points in $\mathcal{S}_{L0}(\ba_i)$. This will result in a mapping with broadcast vectors in $\mathcal{C}_{L1}(\ba_i)$ mapped to adjacent signal points.
\item Repeat steps \ref{map-begn} to \ref{map-end} by considering the remaining effective broadcast vector sets one by one.
\end{enumerate}

For a receiver $R_i$, when we compare the optimal mappings for two different index codes, the code which has less $|\mathcal{S}_L(\ba_i)|$ will perform better as the minimum inter-set distance will be more (note that for both the index codes we do optimal mapping). \\
The mapping algorithm is explained below. Index codes are identified using the corresponding encoding matrices.
\begin{enumerate}
\item The algorithm starts by considering $\mathbb{L}_N$, the set of all index codes of length $N$, for the given ICP. For $R_i$ define,
\begin{equation*}
\eta_i \triangleq \min_{L \in \mathbb{L}_N}{|\mathcal{S}_L(\ba_i)|}
\end{equation*}
\item Find $\eta_1$. If $\eta_1<2^N$ proceed to step \ref{alg-t-nonzero} with $i=1$.
\item If $\eta_1=2^N$, $R_1$ sees the full $2^N$-PSK constellation as the effective broadcast signal set. In such a case all mappings for all the index codes will give same PSK-ICG for $R_1$ with $d^{(1)}_{IS,min}$ same as the minimum Euclidean distance between any two points of $2^N$-PSK constellation. In such a case, any mapping for any index code is optimum for $R_1$. Then consider the next highest priority receiver, $R_2$ and continue until a receiver $R_i$ for which $\eta_i<2^N$ is found. If $\eta_i=2^N$ for all receivers, do an arbitrary mapping and exit. Now consider the case where a receiver $R_i$ for which $\eta_i<2^N$ is found. 
\item \label{alg-t-nonzero} Let $\{L:|\mathcal{S}_L(\ba_i)|=\eta_i\}$ be $\{L_1,L_2,...,L_{n_{L,i}}\}$. For each $L_j, j \in \{1,2,...,{n_{L,i}}\}$, find optimal mappings for $R_i$. Let there be $n_{\mathcal{M},i}$ optimal mappings for each index code and denote the mappings corresponding to index code $L_j$ as $\mathcal{M}_{j1},\mathcal{M}_{j2},...,\mathcal{M}_{jn_{\mathcal{M},i}}$. Define $\mathcal{O}$, the set of ordered pairs as,

\begin{equation*}
\begin{aligned}
\mathcal{O} \triangleq \{&(L_1,\mathcal{M}_{11}),(L_1,\mathcal{M}_{12}),...,(L_1,\mathcal{M}_{1n_{\mathcal{M},i}}),\\
&(L_2,\mathcal{M}_{21}),(L_2,\mathcal{M}_{22}),...,(L_2,\mathcal{M}_{2n_{\mathcal{M},i}}),...,\\
&(L_{n_{L,i}},\mathcal{M}_{{n_{L,i}}1}),(L_{n_{L,i}},\mathcal{M}_{{n_{L,i}}2}),...,\\
&(L_{n_{L,i}},\mathcal{M}_{{n_{L,i}}n_{\mathcal{M},i}})\}.
\end{aligned}
\end{equation*}

The set $\mathcal{O}$ contains all the (index code, mapping) pairs which give the maximum gain possible for $R_i$. Now from this set, identify the pairs which give maximum gain for $R_{i+1}$. For this, choose the pairs which have maximum $d^{(i+1)}_{IS,min}$. Now consider these pairs as set $\mathcal{O}$ and continue until the last receiver $R_m$ is considered and the pairs which have maximum $d^{(m)}_{IS,min}$ is obtained. These are the (index code, mapping) pairs which are optimal.
\end{enumerate}

\subsection{Illustration of Mapping Algorithm}
We illustrate the mapping algorithm given as Algorithm $\ref{alg:a1}$ with an example.
\par Consider the ICP given in Example \ref{exmp:eg1}. Assume that the decreasing order of priority is $(R_1,R_2,R_3,R_4,R_5)$.
Let $N=3$ which is also the length of the optimal index code in this case. Since this is a single unicast ICP, we can find all index codes by considering fitting matrices \cite{Yossef} of rank 3. There are a total of 32 such matrices. For each of these 32 matrices, choose any 3 independent rows as a basis for the row space. So we obtain 32 row spaces (which represents 32 index codes for the given ICP). Of these, only six row spaces are distinct. From Corollary 1 in \cite{Kav2}, the number of index codes possible with the optimal length $c$ for a single-unicast IC problem is given by $\frac{\mu}{c!}\prod^{c-1}_{i=0}(2^c-2^i)$ where $\mu$ is the number of distinct row spaces of c-ranked fitting matrices.
\par For the example under consideration, there are a total of 168 index codes (28 index codes for each distinct row space). So, $\mathbb{L}_3$ contains 168 index codes. $\eta_1 = \min_{L \in \mathbb{L}_3}{|\mathcal{S}_L(\ba_1)|}=4$. There are 84 index codes with $\eta_1=4$ and 32 optimal mappings for $R_1$, for each of these index codes. The set $\mathcal{O}$ has $32*84=2688$ (index code, mapping) pairs which are optimal for $R_1$. One such $(L,\mathcal{M})$ pair has the index code as given in Example \ref{exmp:eg1} and mapping as given in Fig.\ref{figure:fig1}(b).
Consider $R_2$. After all pairs in $\mathcal{O}$ are considered, the maximum value possible for $d^{(2)}_{IS,min}=1.414$ and there are 336 pairs which are optimal for $R_2$. Now consider $R_3$. All 336 pairs gives the same $d^{(3)}_{IS,min}=1.414$. For $R_4$ and $R_5$ all pairs have same minimum inter-set distance and these 336 pairs give the (index code, mapping) pairs which are optimal for the ICP considered. For illustration, four such $(L,\mathcal{M})$ pairs are given below. Index code based on encoding matrix $L$ is given in the form of $(y_1,y_2,y_3)$. $\mathcal{M}$ is given as an ordered list of eight integers, representing the decimal equivalent of the 3-tuple, which is mapped to $(\mathbf{s_1},\mathbf{s_2},...,\mathbf{s_{8}})$ where $(\mathbf{s_1},\mathbf{s_2},...,\mathbf{s_{8}})$ are $2^N$-PSK signal points as shown in Fig. \ref{figure:fig1}.
\begin{itemize}
\item $(\{x_1, x_2+x_3,x_4+x_5\},(0,1,2,3,4,5,6,7))$
\item $(\{x_1, x_2+x_3,x_4+x_5\},(0,1,6,7,4,5,2,3))$
\item $(\{x_1, x_2+x_3,x_1+x_4+x_5\},(0,1,2,3,5,4,7,6))$
\item $(\{x_1, x_1+x_2+x_3,x_4+x_5\},(0,1,2,3,6,7,4,5))$

\end{itemize}

\begin{algorithm}
\caption{Algorithm to find optimal (index code, mapping) pairs for a given ICP.} \label{alg:a1}
\begin{algorithmic}[1]
\State $i\gets 1$
\State Find $\eta_i = \min_{L \in \mathbb{L}_N}{|\mathcal{S}_L(\ba_i)|}$ \label{step2}
\If {($\eta_i=2^N$)}
	\State $i\gets i+1$ \label{step3}
\If {$(i>m)$} 
\State Do an arbitrary mapping and Exit.	
\Else
\State Goto \ref{step2}
\EndIf

\Else
	\begin{itemize}
	\item Consider the set of index codes $\{L_1,L_2,...,L_{n_{L,i}}\}=\{L:|\mathcal{S}_L(\ba_i)|=\eta_i\}$ 
	\item Find $\mathcal{O}$, the set of all (index code, optimal mappings) pairs for $R_i$.
	\end{itemize}
\State $i\gets i+1$ \label{step12}
\If {$(i > m)$} 
\State Output $\mathcal{O}$ and Exit.	
\Else
\State Choose any $(L,\mathcal{M}) \in \mathcal{O}$
\State $\mathcal{O}^{i} \gets \{(L,\mathcal{M})\}$. Find $\delta=d^{(i)}_{IS,min}$.
\State $\mathcal{O} \gets \mathcal{O} \setminus \{(L,\mathcal{M})\}$	 \label{step17}
\If {$(\mathcal{O}=\{\})$}
\State $\mathcal{O} \gets \mathcal{O}^{i}$	
\State Goto \ref{step12}
\Else
\State Consider any $(L,\mathcal{M}) \in \mathcal{O}$. Find $d^{(i)}_{IS,min}$.
\If {($d^{(i)}_{IS,min}>\delta$)}
\State $\mathcal{O}^{i} \gets \{(L,\mathcal{M})\}$, $\delta=d^{(i)}_{IS,min}$. Goto \ref{step17}.
\Else
\If {($d^{(i)}_{IS,min}=\delta$)}
\State $\mathcal{O}^{i} \gets \mathcal{O}^{i} \cup \{(L,\mathcal{M})\}$. Goto \ref{step17}.
\Else
\State Goto \ref{step17}.
\EndIf
\EndIf
\EndIf
\EndIf
\EndIf
\end{algorithmic}
\end{algorithm}

\textit{Claim 1:} Algorithm $\ref{alg:a1}$ guarantees that no other mapping of $2^N$-PSK constellation for any index code, can give PSK-ICG $>g_{1}$ for $R_1$.
\begin{proof}
The coding gain (PSK-ICG) achieved by a receiver is maximized when the minimum inter-set distance is maximum. Consider an index code of length $N$. Using Algorithm $\ref{alg:a1}$, for each of the effective broadcast signal sets of the highest priority receiver, the broadcast vectors in 0-effective broadcast vector set are always mapped to adjacent points. Similarly, the broadcast vectors in 1-effective broadcast vector set are always mapped to adjacent points. These sets of points are placed diametrically opposite to each other. Thus, the minimum inter-set distance is maximized for the chosen index code and the mapping is optimal.
\par When we compare the message error performance of $R_1$ with respect to different possible index codes, the code which has less $|\mathcal{S}_L(\ba_i)|$ performs better. Index codes with minimum $|\mathcal{S}_L(\ba_i)|$ are only considered for mapping in Algorithm $\ref{alg:a1}$. So, the pairs considered by Algorithm $\ref{alg:a1}$ has index codes with minimum $|\mathcal{S}_L(\ba_i)|$ and mappings which are optimal. No other mapping of $2^N$-PSK constellation for any index code of length $N$, can give PSK-ICG $>g_{1}$ for $R_1$. 
\end{proof}

\textit{Claim 2:} Algorithm $\ref{alg:a1}$ guarantees that, any mapping for any index code which gives the PSK-ICG $g_i$ for receivers $R_i$, $i \in \{1,2,...,j-1\}$ cannot give a PSK-ICG  $>g_j$ for $R_j$, $j \leq m$.
\begin{proof}
Algorithm $\ref{alg:a1}$ finds all (index code, mapping) pairs which are optimal for $R_1$. In the next step, among these pairs, which ever gives the maximum gain for $R_2$ are chosen. So, given that $R_1$ has the same PSK-ICG, it is not possible to find another pair for which $R_2$ performs better. Same argument extends to other receivers as well. 
\end{proof}

\par Algorithm $\ref{alg:a1}$ can also be used to obtain optimal (index code, mapping) pairs for a given set of index codes of length $N$. In this case the algorithm must be run by considering the given set of index codes instead of all possible index codes of length $N$. This can be illustrated using the ICP given in Example \ref{exmp:eg2}. Assume that the decreasing order of priority is $(R_1,R_2,R_3,R_4,R_5,R_6)$. Let $N=4$ and assume that only one index code as given in Example \ref{exmp:eg2} need to be considered (the given set of index codes is a singleton set). 
Consider the highest priority receiver $R_1$. Obtain the effective broadcast vector sets seen by $R_1$ for $\ba_1 \in \mathbb{F}^5_2$ and partition these sets. The effective broadcast vector sets and its partitions for $R_1$ are given in Table \ref{table:tbl2}. For any realization of $\bx_{\mathcal{I}_{1}}=\ba_1$ which is not listed in Table \ref{table:tbl2}, the effective broadcast vector set is same as one of the effective broadcast vector sets given in the table.

\begin{table}[ht]
\caption{Effective broadcast vector sets and its partitions (seen by $R_1$) for the IC in Example \ref{exmp:eg2}.}
\centering
\begin{tabular}{c c c c}
\hline
\hline                        
$\ba_1$ & $\mathcal{C}_L(\ba_1)$ & $\mathcal{C}_{L0}(\ba_1)$ & $\mathcal{C}_{L1}(\ba_1)$\\
\hline
$(00000)$ & $\{(0000),(1000)\}$ & $\{(0000)\}$ & $\{(1000)\}$ \\
$(00001)$ & $\{(0001),(1001)\}$ & $\{(0001)\}$ & $\{(1001)\}$ \\
$(00010)$ & $\{(0010),(1010)\}$ & $\{(0010)\}$ & $\{(1010)\}$ \\
$(00011)$ & $\{(0011),(1011)\}$ & $\{(0011)\}$ & $\{(1011)\}$ \\
$(01000)$ & $\{(0100),(1100)\}$ & $\{(0100)\}$ & $\{(1100)\}$ \\
$(01001)$ & $\{(0101),(1101)\}$ & $\{(0101)\}$ & $\{(1101)\}$ \\
$(01010)$ & $\{(0110),(1110)\}$ & $\{(0110)\}$ & $\{(1110)\}$ \\
$(01011)$ & $\{(0111),(1111)\}$ & $\{(0111)\}$ & $\{(1111)\}$ \\
 [1ex]      
\hline
\end{tabular}
\label{table:tbl2}
\end{table}

\begin{figure}
\begin{center}
\includegraphics[clip,scale=0.7]{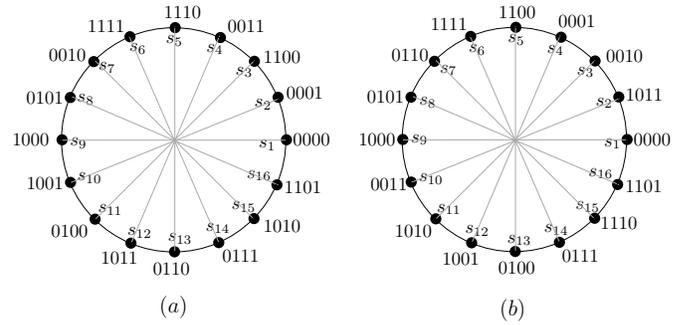} 
\caption{Two 16-PSK mappings for the IC in Example \ref{exmp:eg2}.}
\label{figure:fig2}
\end{center}
\end{figure}

There are $645120$ optimal mappings for $R_1$. The set $\mathcal{O}$ has $645120$ (index code, mapping) pairs which are optimal for $R_1$, with the index code being the same for all the pairs. Consider $R_2$. After all pairs in $\mathcal{O}$ are considered, the maximum value possible for $d^{(2)}_{IS,min}=1.847$ and there are 128 pairs which are optimal for $R_2$. Now consider $R_3$. There are 24 pairs which are optimal with $d^{(3)}_{IS,min}=0.765$. For $R_4$ there are 16 optimal pairs with minimum inter-set distance $d^{(4)}_{IS,min}=0.765$. For $R_5$ and $R_6$ all these pairs give the same minimum inter-set distance. These 16 pairs are the optimal mappings for the IC considered. One of these mappings is given in Fig. \ref{figure:fig2}(a).

\section{Simulation results} \label{sim_section}
\begin{figure}
\includegraphics[clip,scale=0.45]{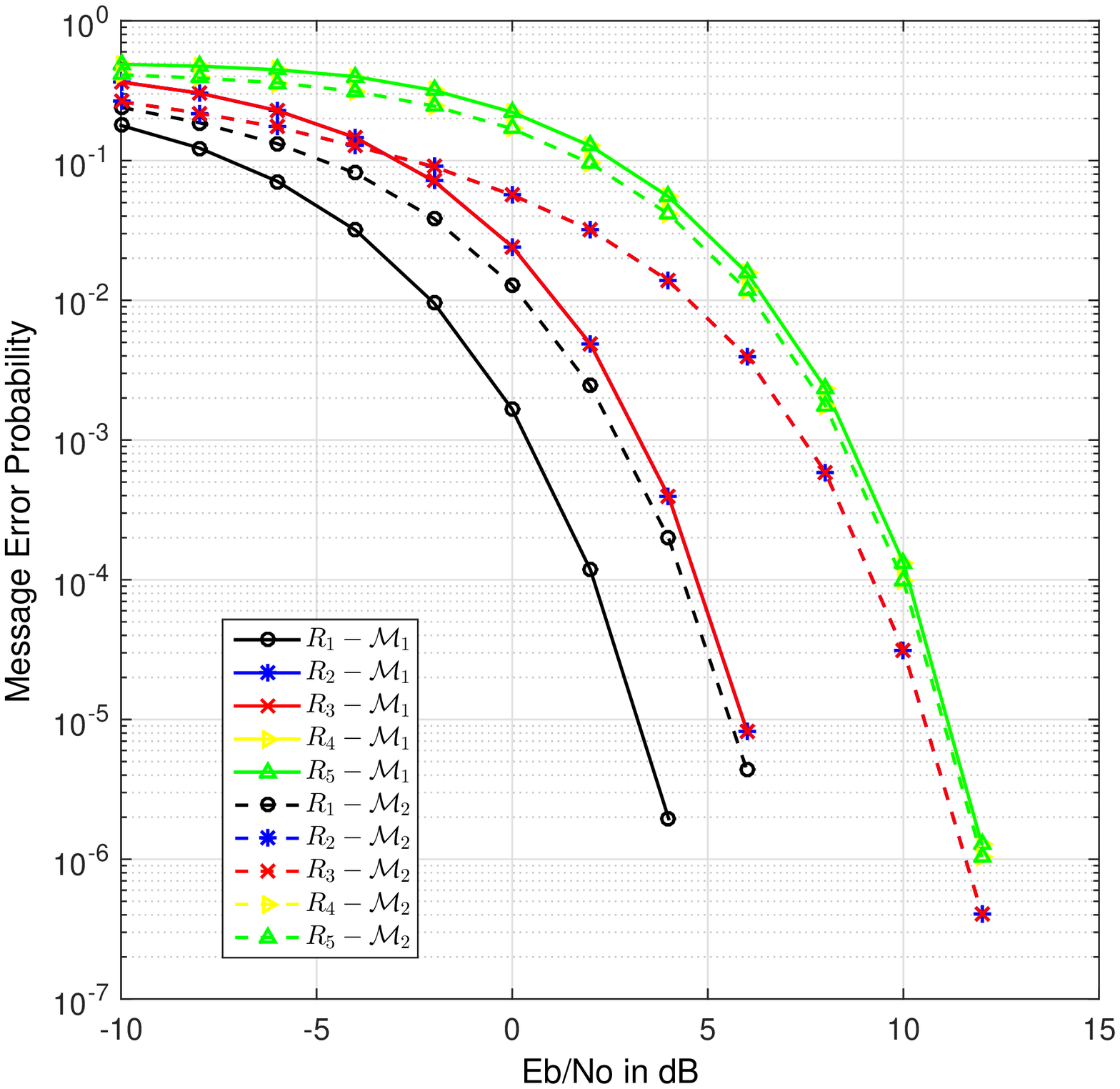} 
\caption{Simulation results comparing the performance of receivers for two different mappings (Example \ref{exmp:eg1}).}
\label{figure:sim1}
\end{figure}

We have considered the ICP given in Example \ref{exmp:eg1} and used Algorithm \ref{alg:a1} to obtain all optimal (index code, mapping) pairs. One such pair, $(L_1,\mathcal{M}_1)$ has the index code  as given in Example \ref{exmp:eg1} and mapping as given in Fig. \ref{figure:fig1}(b). We compared this optimal mapping with another mapping $\mathcal{M}_2$ (shown in Fig. \ref{figure:fig1}(c)) which is not optimal for the same index code, $L_1$. The pair $(L_1,\mathcal{M}_2) \notin \mathcal{O}$, the output set obtained from the execution of the algorithm. We obtained the message error probability of the receivers for the two different mappings, by simulation. The first mapping, ($\mathcal{M}_1$) used Algorithm \ref{alg:a1} and the second mapping ($\mathcal{M}_2$) used an algorithm based on maximizing the minimum Euclidean distances \cite{Anj1}. Simulation results are given in Fig. \ref{figure:sim1}.
  
The performance of receivers $R_1$, $R_2$ and $R_3$ is significantly better with $\mathcal{M}_1$ than with $\mathcal{M}_2$ at high SNR. The minimum inter-set distances are more for $\mathcal{M}_1$ (Fig.\ref{figure:fig1}(b)) than for $\mathcal{M}_2$ (Fig.\ref{figure:fig1}(c)). For receivers $R_4$ and $R_5$, the minimum inter-set distances are same for both the mappings.   

\par
\begin{figure}
\includegraphics[clip,scale=0.45]{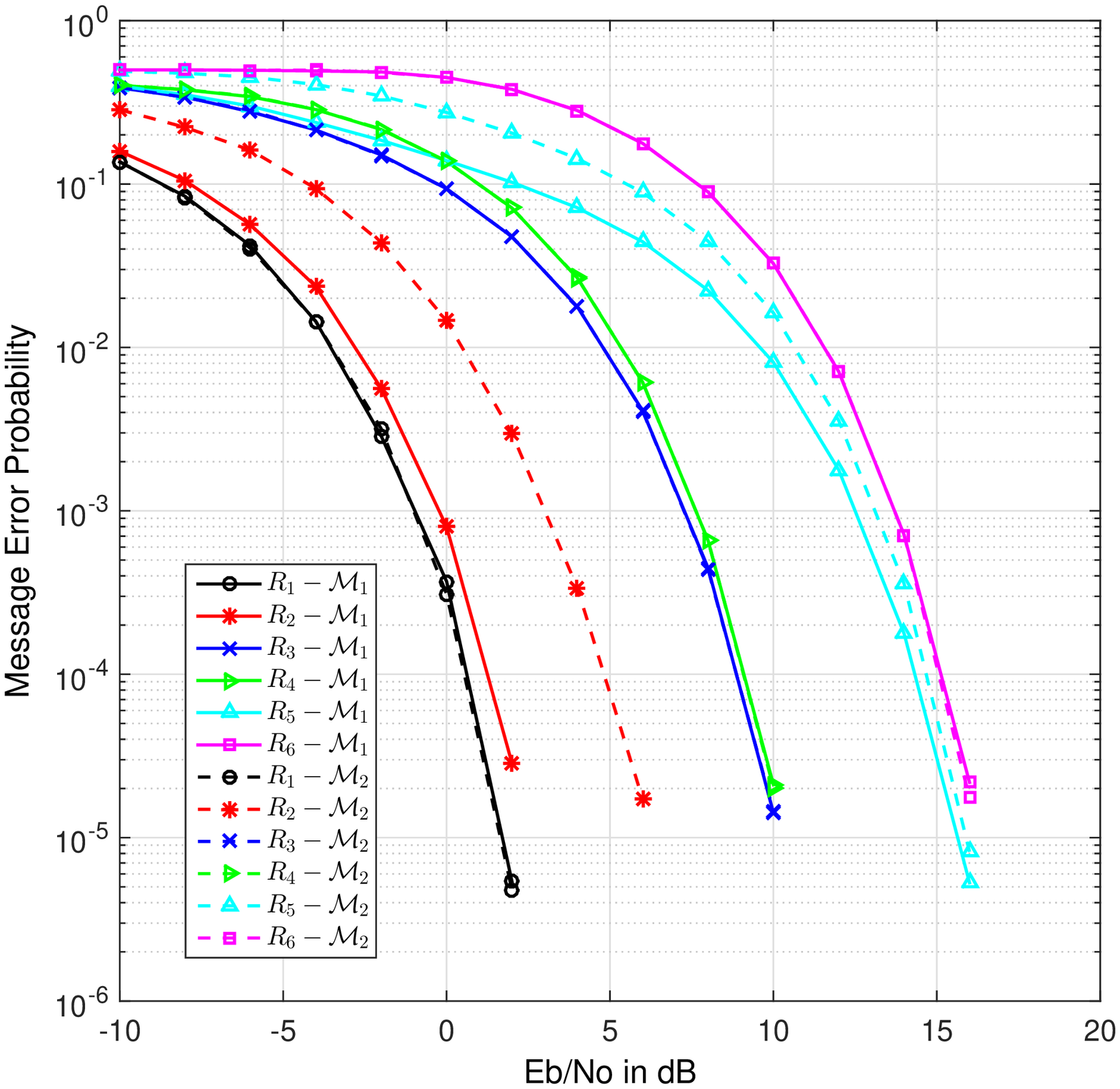} 
\caption{Simulation results comparing the performance of receivers for two different mappings (Example \ref{exmp:eg2}).}
\label{figure:sim2}
\end{figure}

 We have carried out simulation based studies to compare the performance of the receivers for the ICP and the IC given in Example \ref{exmp:eg2} for two different mappings as given in Fig. \ref{figure:fig2}. The mapping ($\mathcal{M}_1$) given in Fig. \ref{figure:fig2}(a) used Algorithm \ref{alg:a1} and the mapping ($\mathcal{M}_2$) given in Fig \ref{figure:fig2}(b) used the algorithm based on maximizing the minimum Euclidean distances \cite{Anj1}. Simulation results are given in Fig. \ref{figure:sim2}.

For $R_1$, $R_3$, $R_4$, $R_5$ and $R_6$, the minimum inter-set distances and hence the performances are the same for both the mappings. But the performance of receiver $R_2$ is significantly better with $\mathcal{M}_1$ than with $\mathcal{M}_2$ at high SNR.
 
\par The simulation results indicate the effectiveness of the algorithm based on minimum inter-set distances (Algorithm \ref{alg:a1}) for mapping the broadcast vectors to PSK signal points. 

It should be noted that Algorithm \ref{alg:a1} does not guarantee that all the receivers will perform better or as good as that with any other algorithm. It is possible that, a mapping based on some algorithm (say, Algorithm 2) gives a better performance to a receiver $R_j$ than that with Algorithm \ref{alg:a1}. But then there will be a receiver $R_i$ which performs better with Algorithm \ref{alg:a1} than with    Algorithm 2, where $R_i$ is a higher priority receiver than $R_j$. In other words, Algorithm \ref{alg:a1} attempts to maximize the gain achieved by the receivers by considering the receivers in the given order of priority. This is further illustrated in Example \ref{exmp:eg3}.

\begin{figure}
\begin{center}
\includegraphics[clip,scale=0.7]{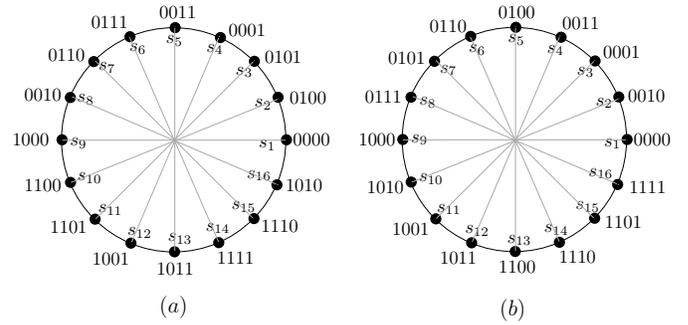} 
\caption{Two 16-PSK mappings for the IC in Example \ref{exmp:eg3}.}
\label{figure:fig3}
\end{center}
\end{figure}

\begin{figure}
\includegraphics[clip,scale=0.45]{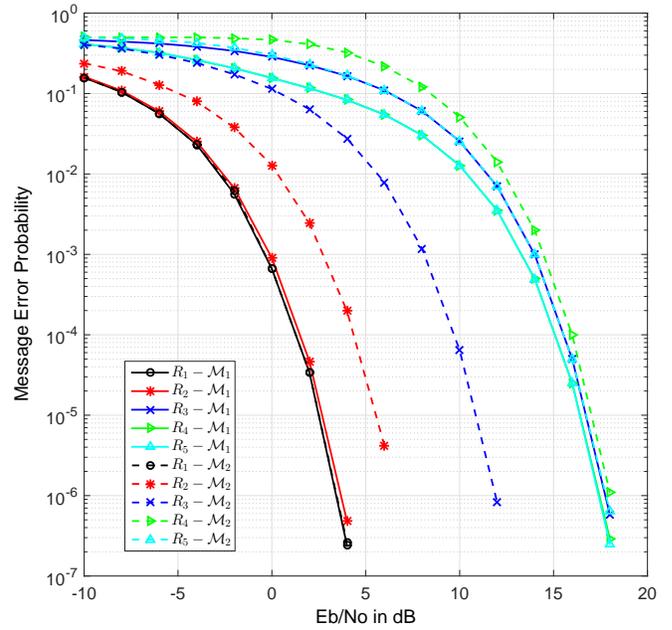} 
\caption{Simulation results comparing the performance of receivers for two different mappings (Example \ref{exmp:eg3}).}
\label{figure:sim3}
\end{figure}

\begin{exmp} \label{exmp:eg3}
Consider the following ICP with $n=m=5$ and $\mathcal{W}_{i}=x_i, ~\forall i \in \{1,2,...,5\}$. The side information available with the receivers is as follows: $\mathcal{K}_1= \{x_2, x_3, x_4, x_5\}, ~\mathcal{K}_2= \{x_1, x_4, x_5\},~ \mathcal{K}_3= \{x_1, x_4\},~\mathcal{K}_4= \{x_2\},~\mathcal{K}_5= \{\}$.
\par For this ICP a scalar linear index code of length $N=4$ (not optimal), is specified as $y_1=x_1+x_2,~ y_2=x_3,~ y_3=x_4,~ y_4=x_5$. Assume that the decreasing order of priority is given as $(R_1,R_2,R_3,R_4,R_5)$. 
\end{exmp}
Using Algorithm \ref{alg:a1}, optimal mappings for the specified IC is obtained, of which one mapping ($\mathcal{M}_1$) is given in Fig. \ref{figure:fig3}(a). Another mapping $\mathcal{M}_2$, is found by using the algorithm based on maximizing the minimum Euclidean distances \cite{Anj1} and is given in Fig. \ref{figure:fig3}(b).
Simulation results comparing the performance of the receivers for these two mappings are given in Fig. \ref{figure:sim3}. 

It is clear from Fig. \ref{figure:sim3} that, $R_2$ performs better with $\mathcal{M}_1$ than with $\mathcal{M}_2$. But $R_3$, which is of lower priority than $R_2$, has better performance with $\mathcal{M}_2$.  

\section{Discussion}
In this paper we have considered index coded PSK modulation and have derived a decision rule for the ML decoder which minimizes the message error probability of the receivers, for a given ICP. We have introduced the concept of inter-set distances and illustrated its importance in noisy index coding problems. It was also shown that at high SNR the dominant factor which decides the message error is the minimum inter-set distance and so an optimal mapping must maximize the minimum inter-set distance.
\par Subsequently, we have considered the problem of finding optimal (index code, mapping) pairs across all possible mappings for all possible index codes of length $N$, for a chosen $2^N$-PSK modulation. This problem was not addressed so far in literature. The algorithm which is proposed for a given ICP, can find (index code, mapping) pairs, each of which gives the best PSK-ICG for the receivers, for any given order of priority.
\par Finding all index codes of a chosen length (greater than or equal to the optimal length) for a given ICP is in general NP hard. If it is too complex to find all the index codes, the algorithm can be executed by considering a chosen set of index codes. But the complexity of the proposed algorithm increases exponentially with the length of the index code.

\section*{Acknowledgment}
This work was supported partly by the Science and Engineering Research Board (SERB) of Department of Science and Technology (DST), Government of India, through J.C. Bose National Fellowship to B. Sundar Rajan.


\begin{thebibliography}{1}
	
	\bibitem{Ahl}
	R. Ahlswede, N. Cai, S. Y. R. Li, and R. W. Yeung, ``Network information flow," {\it{IEEE Trans. Inf. Theory}}, vol. 46, no. 4, pp. 1204-1216, Jul. 2000.	
	
	\bibitem{Birk}
	Y.~Birk and T.~Kol, ``{Coding on demand by an informed source (ISCOD) for efficient broadcast of different supplemental data to caching clients}," {\it{IEEE Trans.Inf.Theory}}, 52(6), June 2006, pp. 2825-2830.
	
	\bibitem{Ong}
	L.Ong and C.K. Ho, ``Optimal Index Codes for a Class of Multicast Networks with Receiver side information," in {\it{Proc. IEEE ICC}}, Ottawa, Canada, June 2012, pp. 2213-2218. 

	\bibitem{Yossef}
	Z. Bar-Yossef, Z. Birk, T.S. Jayaram, and T. Kol, ``{Index coding with side information}," in {\it{Proc. 47th Annu. IEEE symp. Found. Comput. Sci}}, Oct. 2006, pp. 197-206.
	
	\bibitem{Anp1}
	Anoop Thomas, Kavitha Radhakumar, Attada Chandramouli and B. Sundar Rajan, ``Optimal Index Coding with Min-Max Probability of Error over Fading Channels," in {\it{Proc. IEEE PIMRC.,}} Hong Kong, 2015, pp. 889-894
	
	\bibitem{Anp2}
	Anoop Thomas, Kavitha Radhakumar, Attada Chandramouli and B. Sundar Rajan, ``Single Uniprior Index Coding with Min-Max Probability of Error over Fading Channels," Accepted for publication in IEEE Transactions on Vehicular Technology.
	
	\bibitem{Kav1}
	Kavitha Radhakumar and B. Sundar Rajan, ``On the number of optimal index codes," Proceedings of IEEE International Symposium on Information Theory, (ISIT 2015), Hong Kong, 14-19 June 2015, pp. 1044-1048.

	\bibitem{Kav2}
	Kavitha Radhakumar, Niranjana Ambadi and B. Sundar Rajan, ``On the Number of Optimal Linear Index Codes for Unicast Index Coding Problems," in {\it{Proc. 47th Annu. IEEE Wireless Communications and Networking Conference}}, Doha, Qatar, April 2016, pp. 1897-1903.
		
	
	\bibitem{Lpn}
	L. Natarajan, Y. Hong and E. Viterbo, ``Index codes for the Gaussian Broadcast Channel using Quadrature Amplitude Modulation," in {\it{IEEE Commun. Lett.}}, August 2015, pp. 1291-1294. 
	
	\bibitem{Anj1}
	Anjana A. Mahesh and B. Sundar Rajan, ``Index coded PSK Modulation," in {\it{Proc. 47th Annu. IEEE Wireless Communications and Networking Conference}}, Doha, Qatar, April 2016, pp. 1890-1896. 

	\bibitem{Vit}
	A. J. Viterbi and J. K. Omura, {\it{Principles of Digital Communication and Coding}}, New York: Dover Publications, 2009, pp. 47-64.  
	
\end{thebibliography}
\end{document}